\documentclass[a4paper,12pt]{article}

\usepackage{graphicx}
\usepackage{amsfonts, amsmath}
\usepackage{amssymb}
\usepackage{mathrsfs}
\usepackage{setspace}
\usepackage[latin1]{inputenc}
\usepackage{comment}
\usepackage{cite}
\usepackage{slashed}
\usepackage{bbold}
\usepackage{subfig}
\usepackage{hyperref}

\DeclareMathOperator{\Tr}{Tr}

\title{\vspace{-3.5cm}\begin{flushright}
 \small SU-4252-919\\ IMSc/2011/9/10 \\
\end{flushright}
\vspace{3.75cm} Quantum Gravity: Mixed States from Diffeomorphism
Anomalies}

\author{\textbf{A. P. Balachandran}\footnote{bal@phy.syr.edu} \\
\emph{Department of Physics, Syracuse University,} \\
\emph{Syracuse, NY 13244-1130, USA} \\ and \\ \emph{International Institute
of
Physics (IIP-UFRN)} \\
\emph{Av. Odilon Gomes de Lima 1722, 59078-400 Natal, Brazil}\vspace*{1cm}
\\
\textbf{Amilcar R. de Queiroz}\footnote{amilcarq@unb.br} \\ \emph{Instituto
de
Fisica, Universidade de
Brasilia,} \\ \emph{Caixa Postal 04455, 70919-970, Brasilia, DF, Brazil}}

\begin{document}

\maketitle

\begin{abstract}
In a previous paper, we discussed simple examples like particle on a circle
and
molecules to argue that mixed states can arise from anomalous symmetries.
This
idea was applied to the breakdown (anomaly) of color $SU(3)$ in the
presence
of non-abelian monopoles. Such mixed states create entropy as well. 

In this article, we extend these ideas to the topological geons
of Friedman and Sorkin in quantum gravity. The ``large diffeos'' or mapping
class groups can become
anomalous in their quantum theory as we show. One way to eliminate
these anomalies is to use mixed states, thereby creating entropy. These
ideas
may have something to do with black hole entropy as we speculate. 
\end{abstract}

\section{Introduction}

Diffeomorphisms of space-time play the role of gauge transformations in
gravitational theories. Just as gauge invariance is basic in gauge
theories, so
too is diffeomorphism (diffeo) invariance in gravity theories.

Diffeos can become anomalous on quantization of gravity models. If that
happens,
these models cannot serve as descriptions of quantum gravitating systems.

There have been several investigations of diffeo anomalies in models of
quantum
gravity with matter in the past. For example, Alvarez-Gaumé and Witten
showed
the absence of these anomalies in string theories
\cite{AlvarezGaume1984269}.

But all these studies have dealt with ``small diffeos'' or the identity
component of the diffeo group. Some references that deals with anomalies
associated with ``large diffeos'' are
\cite{witten-global-1985,Surya:1997ej}. For ``large gauge''
anomalies, see
\cite{Witten1982324,Balachandran:1996jz,Balachandran:1996iv,
Balachandran:1998zq}.

We here consider asymptotically flat space-times and focus on the diffeo
group
$D^\infty$ of the spatial slice $M^d$ of dimension $d$ which keeps the flat
metric and a frame at $\infty$ fixed. The identity component $D_0^\infty$
of
$D^\infty$ is what is required to act trivially on quantum states by the
diffeomorphism constraints. The group of ``large diffeos''
$D^\infty/D^\infty_0$
is called the mapping class group of $M^d$. It can act non-trivially on
quantum
states, but observables commute with it. It is a discrete group.

If $M^d$ is $\mathbb{R}^d$, then $D^\infty/D^\infty_0=\{e\}$. So we need
more
interesting models of $M^d$ for work on large spatial diffeos. They are
provided
by the geon manifolds of Friedman and Sorkin
\cite{PhysRevLett.44.1100,PhysRevLett.45.148,Aneziris:1989cr}. We study such manifolds
for $d=2$ and $d=3$, and show that they can become anomalous.

Our approach towards this demonstration is based on the work of Esteve
\cite{PhysRevD.34.674,aguado2001vacuum,PhysRevD.34.674}.
It is as follows. We consider the Dirac operator or the Laplacian of a
matter field on the manifold $M^d$. The eigenmodes of these operators enter
the
mode expansion of the matter field and its second quantization. The Dirac
operator or the Laplacian must be self-adjoint in order to have  a
complete set of orthornormal eigenstates and real eigenvalues so that it
can be
used in the above mode expansion.

Now the proper definition of these operators as self-adjoint operators
involves the choice of a domain for them: there can be several inequivalent
choices
leading to different physics. We then show cases where these domains are
changed
by $D^\infty/D^\infty_0$ proving that they are anomalous
\cite{PhysRevD.34.674}. These
anomalies are similar to the parity and time-reversal anomalies for certain
domains of the Laplacian for a particle on a circle or for several
different
molecules \cite{Balachandran:1996jz,Balachandran:2011bv}.

The large diffeos leave the Dirac operator and the Laplacian
invariant if the
metric is also transformed. Still the domain can be changed by ``large diffeos'' so that ``a
classical
diffeo symmetry becomes anomalous in quantum theory''.

There is another approach to this question of diffeos and their
compatibility
with domains. It is based on Witten's proof of positive energy theorem
\cite{Witten1981}.
Witten uses a Dirac operator on a spatial slice which includes the
influence of
both gravity and matter. Its definition involves the proper choice of
domains
to guarantee ellipticity \cite{Witten1981,Gibbons1983enerpos}. If diffeos change this
domain, then they are surely anomalous.

But this line of argument requires more articulation as we will see.

In section 2, we give a presentation of simple geon manifolds $M^d$ for
$d=2$
and $d=3$ in a form convenient for the present work. 

In section 3, we discuss domains for Dirac operators where they are
self-adjoint. We also discuss how we can recover spatial topology from
these
domains, thereby making progress in the problem of the confused quantum
baby
\cite{Balachandran2008f}.

In section 4, we study the action of the diffeo group on domains and show
how it
can change them. Impure states which eliminate such anomalies are also
constructed.

In section 5, we look at the Dirac operator used by Witten in his proof of
the
positive energy theorem, but this time on geon manifolds. It includes
gravity,
indeed the
ADM Hamiltonian can be written using it. Its proper treatment certainly
involves
a domain choice. Just as previously, large diffeos change several of these
domains.

In the concluding section 6, we speculate on the entropy of the mixed states associated with large
diffeos and why it may have something to do with black hole
entropy.

\section{A Presentation of Geon Manifolds}

If $M_1$ and $M_2$ are two manifolds of dimension $d$, their connected sum
$\#$
is defined as follows \cite{balachandran1991classical}. Remove balls $B_i$, with
$i=1,2$, from $M_i$. Then $M_i\backslash B_i$ are manifolds with spheres
$S^{d-1}$ as boundaries. Identify these spheres to obtain $M_1\# M_2$. If
$M_i$
are oriented, and this identification is done with orientation reversal,
then
$M_1\# M_2$ is also oriented.

There is a class of closed (that is, compact and boundaryless) manifolds
for
$d=2$ and $d=3$ which are called primes. \emph{All} closed (that is,
compact and
boundaryless) manifolds are
connected sums of primes.

For $d=2$, there is only one orientable prime, namely the two-torus $T^2$.
For
$d=3$, there are an infinity of them , one being the three-torus $T^3$.

In this paper, we will focus on $\mathbb{R}^d\# T^d$, with $d=2,3$, because
they
are relatively simple and also illustrate our ideas. They are
asymptotically
flat: all asymptotically flat two- and three-dimensional manifolds with one
asymptotic region are obtained by attaching a finite number of primes $P_\alpha$ of dimension
$d$
to $\mathbb{R}^d$, that is, they are $\mathbb{R}^d\# P_1 \# P_2 \# ... \#
P_k$.

Now, there is an elegant way to present $\mathbb{R}^d\# P$ if $P$ is a
prime. We
take out a ball $B^d$ from $\mathbb{R}^d$ to obtain $\mathbb{R}^d\backslash B^d$ and then make suitable
identifications
of points on the boundary $\partial\left(\mathbb{R}^d\backslash B^d
\right)$ of
$\mathbb{R}^d\backslash B^d$.

\begin{figure}[!ht]
  \begin{center}
         \includegraphics[scale=0.3,angle=-90]{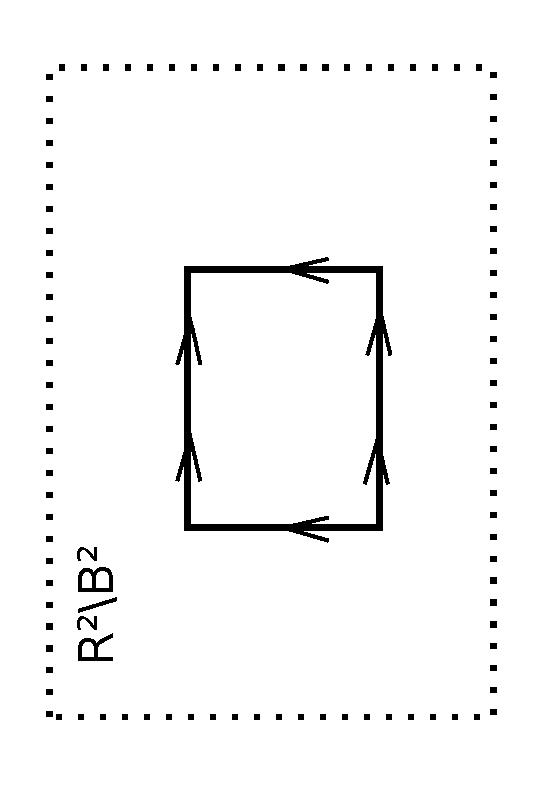}
          \end{center}
  \caption{A presentation of $\mathbb{R}^2\# T^2$}
  \label{fig1}
\end{figure}

Let us show this for $\mathbb{R}^2\# T^2$. A square with its interior is as
good
as a two-ball (disk), as the former can be deformed to the latter. So we
remove
such a square, call it $B^2$, from $\mathbb{R}^2$. Then we identify
opposite
sides
of $\partial\left(\mathbb{R}^2\backslash B^2 \right)$ to obtain
$\mathbb{R}^2\#
T^2$. Figure \ref{fig1} displays this construction.

A similar method works for $\mathbb{R}^3\# T^3$: we remove a cube from
$\mathbb{R}^3$ and identify opposite faces.

We can think of the empty square and cube with their identifications in
$\mathbb{R}^d$, with $d=2,3$, as the spatial location of geons where we
carved
holes in $\mathbb{R}^d$ are where the geons have been attached.

This presentation is very convenient since we can use the flat metrics away
from
the holes to treat the Dirac operator and Laplacian on
$\mathbb{R}^d\backslash
B^d$, $B^d$ being the square and the cube for $d=2,3$. The identifications
at the
boundaries can be incorporated in the operator domains. That gives us
relatively
simple models to deal with. But we do not expect the conclusion to be
sensitive
to our choice of models.

\section{On Domains of Operators}

\subsection{$\mathbb{R}^2\# T^2$}

 The Dirac operator we deal with is 
 \begin{align}
      \label{Dirac-op-1}
       i\slashed{D} &= i\gamma\cdot \partial \\
       \gamma^a &= \sigma^a, ~~~~ a=1,2 \\
       \partial_a &= \frac{\partial}{\partial x^a},
 \end{align}
where $(x^1,x^2)$ are the Cartesian coordinates on $\mathbb{R}^2\backslash
B^2$
and $\sigma^a$ are Pauli matrices. Gauge fields can be included in
(\ref{Dirac-op-1}), but we do not do so here for simplicity. See below for
further
comments on gauge fields.

The operator $i\slashed{D}$ is defined on $\mathbb{R}^2\backslash B^2$
which has
the square as boundary.

The boundary conditions on $\partial(\mathbb{R}^2\backslash B^2)$ define
the domain $\mathcal{D}$ of $i\slashed{D}$. It must be chosen so that $i\slashed{D}$ is self-adjoint. For technical details about the latter, see \cite{atiyah1975spectral,atiyah1975spectralb,atiyah1976spectral}.

The topology of the underlying manifold is also encoded in $\mathcal{D}$. This
aspect is of importance for this work. Thus if the underlying manifold has
to
have
the topology of $T^2$, then $\mathcal{D}$ must be a representation space for
continuous functions on $T^2$. That is to say, it must be a module for the
$\mathbb{C}^*$-algebra $\mathcal{C}^0(T^2)$. Then we can recover $T^2$ as a
topological space by the Gelfand-Naimark theorem \cite{gelfand1943inclusion}.

If we want a more refined statement on $T^2$ and recover also its
differential
structure, that can be done by requiring that $\mathcal{D}$ is a module for the
algebra of once-differentiable functions $\mathcal{C}^1(T^2)$ on $T^2$. We
can
keep on going in this manner and require that $\mathcal{D}$ is a module  for
$\mathcal{C}^\infty(T^2)$, the algebra of infinitely differentiable
functions on
$T^2$.

We will see that such a domain exists. It is one where $\slashed{D}^k$ is
``essentially self-adjoint'' for all $k$.

\begin{figure}[!ht]
  \begin{center}
         \includegraphics[scale=0.3,angle=-90]{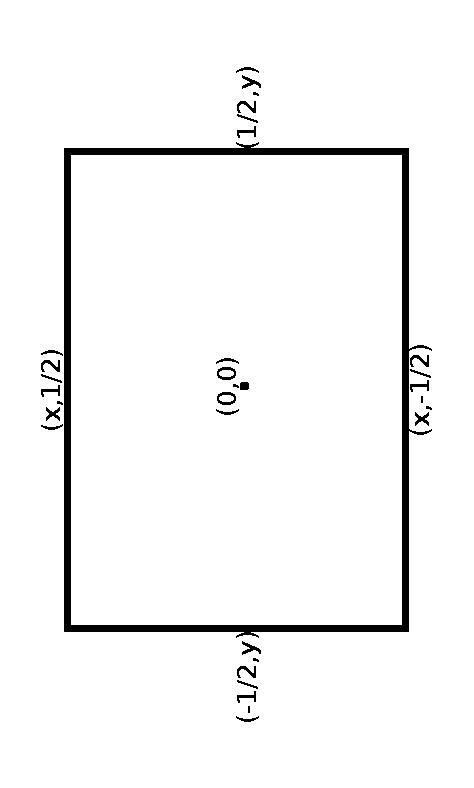}
          \end{center}
  \caption{The coordinates for $\partial(\mathbb{R}^2\backslash B^2)$}
  \label{fig2}
\end{figure}

Let us see how to find such a domain. For purposes of describing it, let us
choose an origin of $\mathbb{R}^2$ in the middle of the square, and give
its
boundaries the coordinates
\begin{align}
      \label{bound-cond-1}
      \left\{ (\pm\frac{1}{2},y) \right. &: \left. -\frac{1}{2}\leq y \leq
\frac{1}{2} \right\}, ~~~ \textrm{and} \\
      \left\{ (x,\pm\frac{1}{2}) \right. &: \left. -\frac{1}{2}\leq x \leq
\frac{1}{2} \right\}.
\end{align}
See Figure \ref{fig2}.

Then in $\mathcal{D}$, call it $\mathcal{D}^{(0)}$, we allow only smooth
$L^2$-functions in $\mathbb{R}^2\backslash B^2$ which fulfill the boundary
conditions
\begin{align}
      \label{bound-cond-2}
       \mathcal{D}^{(0)}=\left\{\psi: \psi(x,\frac{1}{2})=e^{i\theta_1}
\psi(x,-\frac{1}{2}), \psi(\frac{1}{2},y)=e^{i\theta_2}
\psi(-\frac{1}{2},y),
\theta_i\in\mathbb{R}\right\}
\end{align}

Functions $\chi$ on $\mathcal{C}^0(T^2)$ are periodic on the square:
$\chi(x,\frac{1}{2})=\chi(x,-\frac{1}{2})$ and 
$\chi(\frac{1}{2},y)=\chi(-\frac{1}{2},y)$. This is in accordance with
Figure
(\ref{fig1}) for $e^{i\theta_i}=1$. One sees that if $\psi$ fulfills
(\ref{bound-cond-2}), so does $\chi\psi$. As we have taken $\mathcal{D}^{(0)}$ to
consist of smooth functions, we should take $\chi$'s also to be smooth.
Then
the completion of the algebra of such $\chi$'s in the sup norm gives us
back
$\mathcal{C}^0(T^2)$.

It is not difficult to prove that $\slashed{D}$ is essentially self-adjoint
on
$\mathcal{D}^{(0)}$

But $\slashed{D}^{2}$ is not self-adjoint for domain $\mathcal{D}^{(0)}$. For
that, we require the domain
\begin{align}
       \mathcal{D}^{(1)}=\left\{\psi\in \mathcal{D}^{(0)}:
\frac{\partial\psi}{\partial y}(x,\frac{1}{2}) =
e^{i\theta_1}\frac{\partial
\psi}{\partial y}(x,-\frac{1}{2}), \right. \\
      \left. \frac{\partial\psi}{\partial x}(\frac{1}{2},y) =
e^{i\theta_2}\frac{\partial \psi}{\partial x}(-\frac{1}{2},y)\right\}
\end{align}
That is, both $\psi$ \emph{and} its normal derivatives to the boundary must
be
quasi-periodic.

For self-adjointness of $\slashed{D}^N$, we similarly require a domain
\begin{align}
       \mathcal{D}^{(N-1)}=\left\{\psi\in \mathcal{D}^{(0)}:
\frac{\partial^k\psi}{\partial y^k}(x,\frac{1}{2}) =
e^{i\theta_1}\frac{\partial^k \psi}{\partial y^k}(x,-\frac{1}{2}), \right.
\\
      \left. \frac{\partial^k\psi}{\partial x^k}(\frac{1}{2},y) =
e^{i\theta_2}\frac{\partial^k \psi}{\partial x^k}(-\frac{1}{2},y), ~
      \forall k <\leq N-1 \right\}.
\end{align}

On $\mathcal{D}^{(N-1)}$, $\slashed{D}^N$ is self-adjoint whereas $\slashed{D}^m$
for $m<N$ are only essentially self-adjoint. But that is enough for us.

For a domain $\mathcal{D}^{(\infty)}$ which is a module for
$\mathcal{C}^\infty(T^2)$, we should take the intersection of all
$\mathcal{D}^{(N)}$, that is,
\begin{equation}
       \mathcal{D}^{(\infty)}=\bigcap_{N}  \mathcal{D}^{(N)}.
\end{equation}
On $\mathcal{D}^{(\infty)}$, $\slashed{D}^N$ are essentially self-adjoint for all
$N$.

\begin{figure}[!ht]
  \begin{center}
         \includegraphics[scale=0.3,angle=-90]{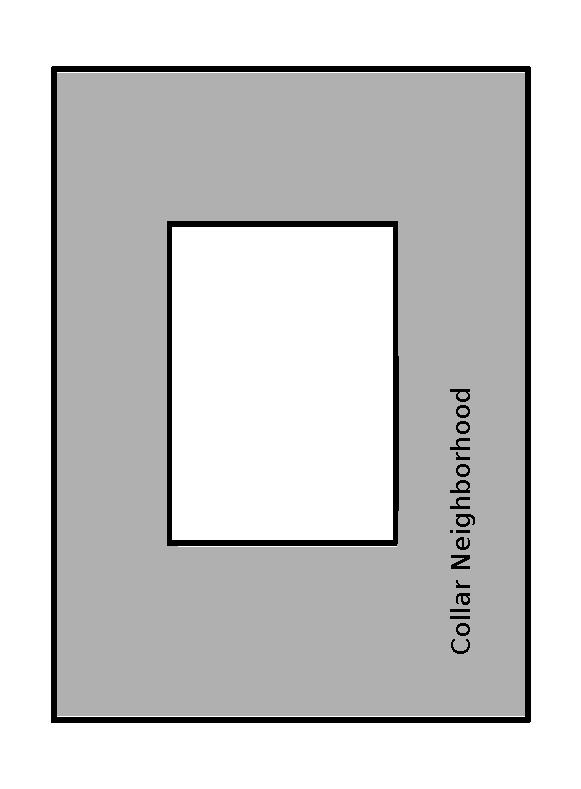}
          \end{center}
  \caption{The collar neighborhood of $\partial(\mathbb{R}^2\backslash
B^2)$}
  \label{fig3}
\end{figure}

It is easy to see that $\mathcal{D}^{(\infty)}$ exists. For that, consider smooth
functions in $\mathcal{D}^{(0)}$ which are constant on a ``collar neighborhood''
of $\partial(\mathbb{R}^2\backslash B^2)$. This neighborhood is shaded in
Figure
(\ref{fig3}). These functions belong to $\mathcal{D}^{(\infty)}$.

\subsection{Absorbing Boundary Conditions into Gauge Fields}

The boundary data of (\ref{bound-cond-2}) can be described in terms of a
$U(1)$-valued field
$U$ on $\partial(\mathbb{R}^2\backslash B^2)$ as we now show. The definition of $U$ is
\begin{align}
      U(x,+\frac{1}{2}) &= e^{i\theta_1} U(x,-\frac{1}{2}), \\
      U(+\frac{1}{2},y) &= e^{i\theta_2} U(-\frac{1}{2},y),
\end{align}
with $|x|,|y|\leq 1/2$.

If the domain $\mathcal{D}^{(0)}$ is called $\mathcal{D}^{(0)}_U$, then the fields
in $\mathcal{D}^{(0)}_{\mathbb{1}}$ \footnote{Here $\mathbb{1}$ has the constant
value $1$ on
$\partial(\mathbb{R}^2\backslash B^2)$.} have
$e^{i\theta_i}=1$. Also if
\begin{equation}
      \chi_1\in \mathcal{D}^{(0)}_\mathbb{1},
\end{equation}
then
\begin{equation}
      U \chi_1|_{\partial(\mathbb{R}^2\backslash B^2)}
\end{equation}
behaves on the boundary according to (\ref{bound-cond-2}).

\begin{figure}[!ht]
  \begin{center}
         \includegraphics[scale=0.3,angle=-90]{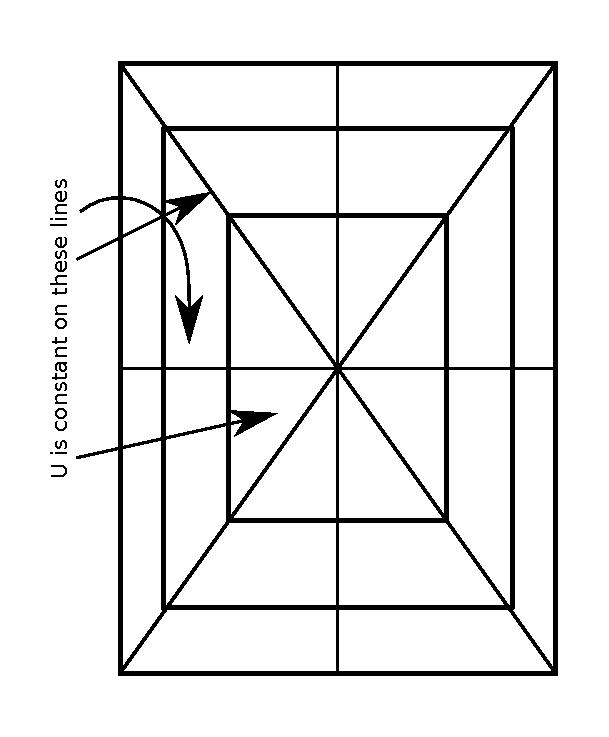}
          \end{center}
  \caption{Extension of $U$ to $\mathbb{R}^2\backslash B^2$ from
$\partial(\mathbb{R}^2\backslash B^2)$.}
  \label{fig4}
\end{figure}

We can extend $U$ to all of $\mathbb{R}^2\backslash B^2$ easily. For
example, we
can cover this space by squares and declare $U$ to be constant on radial
lines
as in Figure \ref{fig4}. Across each square, it is quasi-periodic as in the square of Figure 2.

But this extended $U$ is not differentiable at the corners of the squares.

\begin{figure}[!ht]
  \begin{center}
         \includegraphics[scale=0.4,angle=-90]{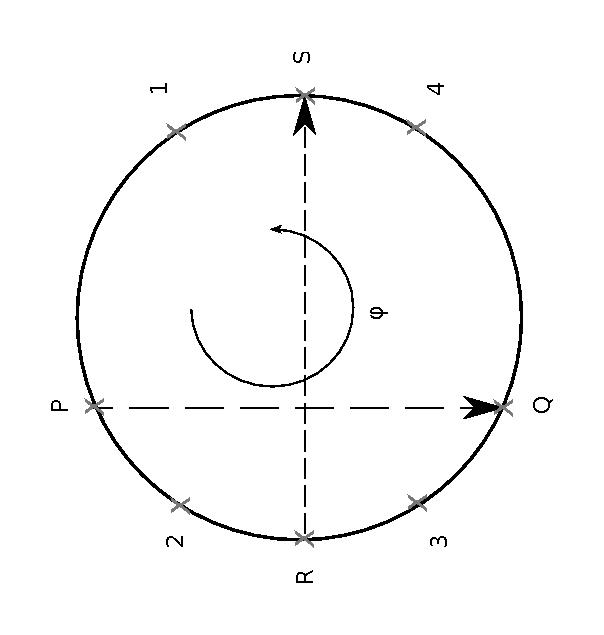}
          \end{center}
  \caption{A smooth $V$ on $S^1$:  $V(S)=e^{i\theta_1(\varphi(P))} V(R)$ ($V(P)=e^{i\theta_2(\varphi(Q))} V(Q)$), where $P\in [1,2]$ and $Q\in [3,4]$
($R\in [2,3]$, $S\in[4,1]$) are vertically (horizontally) opposite points. The
functions $e^{i\theta_i(\varphi)}$, with $i=1,2$, are smooth functions on
$S^1$, with $\varphi$ being the angular coordinate on $S^1$. At point $S\in S^1$, this coordinate is $\varphi(S)$. Similarly at point $P$, it is $\varphi(P)$, and so on.}
  \label{fig5}
\end{figure}

But there is a different domain with a different $V$ at the boundary where this problem can be overcome. Instead of a square for $B^2$, let us choose a disk with boundary as a smooth circle $S^1$ as in Figure \ref{fig5}.

For such $V$ explicitly defined in Figure \ref{fig5}, we can define a new domain
\begin{equation}
       \mathcal{D}^{(0)}_V=\left\{ \psi: \psi|_{\partial(\mathbb{R}^2\backslash
B^2)}=V\chi_1|_{\partial(\mathbb{R}^2\backslash B^2)}  \right\}
\end{equation}
where $\chi_1\in  \mathcal{D}^{(0)}_\mathbb{1}$.

The operator $i\slashed{D}$ is (essentially) self-adjoint on
$\mathcal{D}^{(0)}_V$. Also $\mathcal{D}^{(0)}_V$ is a module for
$\mathcal{C}^\infty(T^2)$ just as we want.

As $V$ is smooth on $\partial(\mathbb{R}^2\backslash B^2)$, there is no
problem in extending it as a smooth $U(1)$-valued function $V$ on all of
$\mathbb{R}^2\backslash B^2$. We may want to require that $V$ approaches
the
same constant value $V_\infty$ as $|x|\to\infty$ in any direction., but
that too
is easily arranged.

Now if $i\slashed{D}$ has domain $\mathcal{D}^{(0)}_U$, then 
\begin{equation}
      \label{dirac-2}
      V^{-1} i\slashed{D} V
\end{equation}
has domain $\mathcal{D}^{(0)}_\mathbb{1}$ as $\mathcal{D}^{(0)}_U$ is determined
only by the boundary value of $V$.

The operator (\ref{dirac-2}) is a Dirac operator with a flat connection
$V^{-1} (
D V)$, that is,
\begin{equation}
      V^{-1} i\slashed{D} V=i\gamma\cdot (D+ V^{-1} (D V)).
\end{equation}

Thus for such a $U$, we can work with a fixed domain $\mathcal{D}^{(0)}_\mathbb{1}$
and a Dirac operator with a connection.

We have earlier discussed this transformation of boundary conditions to
connections in the simple case of a particle on a circle
\cite{Balachandran1995c}.

\subsection{The Role of the $\pi_1$-Group}

\begin{figure}
  \centering
  \subfloat[$\mathbb{R}^2\#
T^2$]{\label{fig:gull}\includegraphics[scale=0.25,angle=-90]{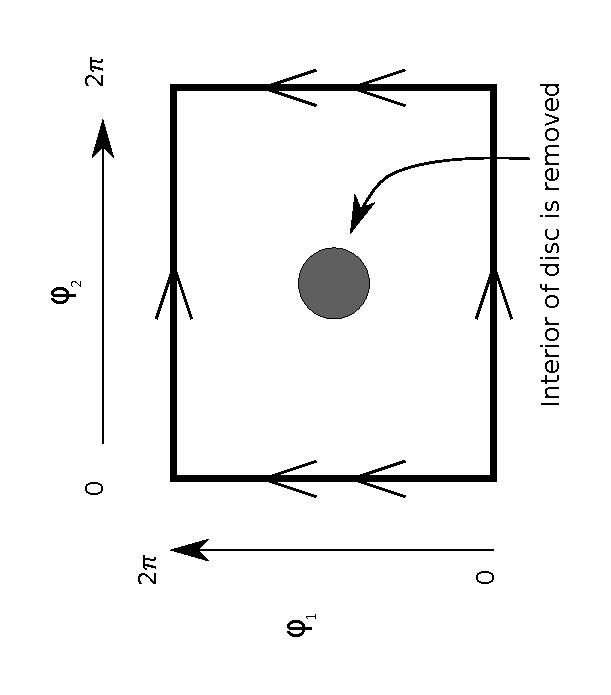}}              
  \subfloat[$a,b,c$
cycles]{\label{fig:tiger}\includegraphics[scale=0.25,angle=-90]{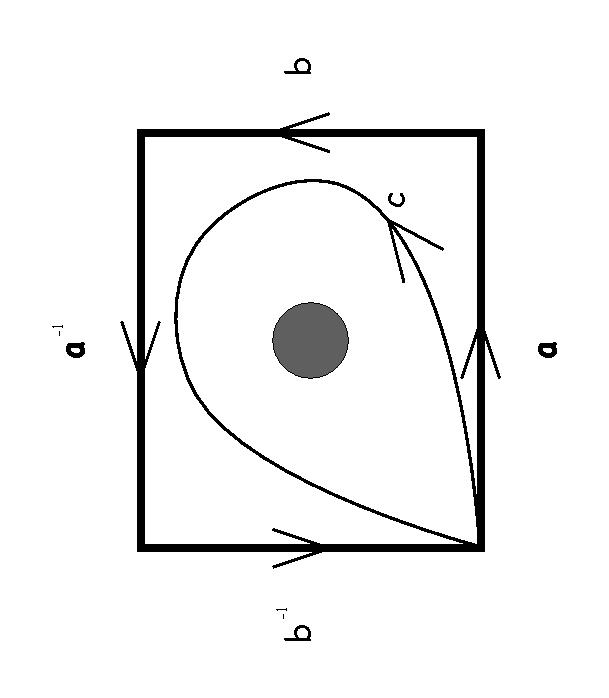}}
   \caption{(a) A new presentation of $\mathbb{R}^2\# T^2$; (b) The $a,b,c$
cycles of $\pi_1(\mathbb{R}^2\# T^2)$.}
  \label{fig6}
\end{figure}

The above discussion can be framed differently \cite{balachandran1991classical}. The
manifold
$\mathbb{R}^2 \# T^2$ is multiply connected. Its fundamental group
$\pi_1(\mathbb{R}^2 \# T^2)$ is actually nonabelian with presentation 
\begin{equation}
      \pi_1(\mathbb{R}^2 \# T^2)=\langle a,b,c:
c=aba^{-1}b^{-1},ac=ca,bc=cb\rangle.
\end{equation}
See the paper \cite{Aneziris:1989cr} for a proof. For $T^2$,
$c=e$, the identity. But here $c$ generates the non-trivial center of the
fundamental group. It comes from the fact that a loop ``circling completely
the
geon'' cannot be deformed to a point. Figure \ref{fig6} shows the $a,b$
and
$c$ cycles.

On multiply connected spaces such as $\mathbb{R}^2\# T^2$, there is a
vector
bundle with a flat connection which implements a unitary irreducible
representation (UIRR) of $\pi_1(\mathbb{R} \#
T^2)$. Its sections then go to define domains of operators like
$i\slashed{D}$ \cite{balachandran1991classical}.

In the example above, we chose a UIRR wherein
\begin{align}
      a &\to e^{i\theta_1}\mathbb{1}, \\
      b &\to e^{i\theta_2}\mathbb{1},
\end{align}
so that
\begin{equation}
      c \to \mathbb{1},
\end{equation}
and the representation is abelian.

There are non-abelian representations as well. A simple example is provided
by the ``rational torus''. In the $N\times N$ representation of $\pi_1(\mathbb{R}^2 \# T^2)$
by
the
rational torus, $a,b$ are represented by ``clock''- and ``shift''- operators $U_1$
and
$U_2$, that is,
\begin{align}
      a &\to U_1, \\
      b &\to U_2,
\end{align}
while $c$ is a root of unity
\begin{equation}
      \label{root-unity-c}
      c=e^{i\frac{2\pi}{N}} \mathbb{1}_{N\times N}.
\end{equation}
We also impose the conditions
\begin{align}
      U_1^N &=e^{i\theta_1}\mathbb{1}, \\
      U_2^N &=e^{i\theta_2}\mathbb{1}.
\end{align}

As a side remark, we observe that one can also represent $c$ by
$e^{i2\pi\frac{p}{N}}\mathbb{1}_{N\times N}$ where $p$ is a fixed integer
in
$[1,2,...,N-1]$. If this differs from (\ref{root-unity-c}), then it is a
new
representation. 

How do we adopt this representation to define domains for Dirac operators?

That is really easy. For $\mathbb{R}^2\# T^2$ we let $i\slashed{D}$ to act
on
$\mathbb{C}^N$-valued spinors $\psi$ so that it has spinor and ``flavor''
indices $a(\in[1,2])$ and $\rho(\in[1,2,...,N])$, respectively. That is, $\psi=\left(
\psi^{a,\rho}\right)$. Then on $\partial(\mathbb{R}^2\backslash B^2)$, we
replace $e^{i\theta_i}$ in (\ref{bound-cond-2}) by $U_i$, where the
matrices
$U_i$ act on the flavor index $\rho$.

\subsection{Generalizations}

\begin{figure}[!ht]
  \begin{center}
         \includegraphics[scale=0.3,angle=-90]{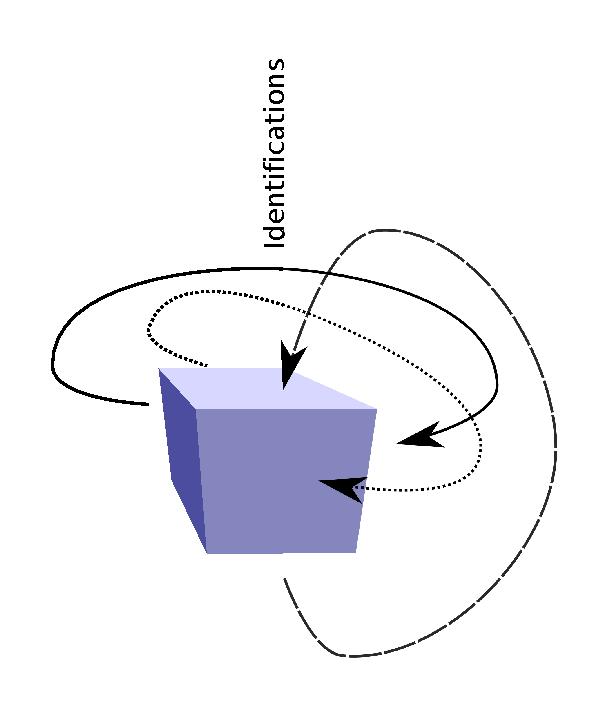}
          \end{center}
  \caption{Opposite faces of the cube should be identified. The figure
shows the
identified top- and bottom-faces. Similar identifications must be made on
side
faces.}
  \label{fig7}
\end{figure}

In three spatial dimensions as well, $\mathbb{R}^3\# P$, where $P$ is an
oriented prime, can be represented by carving out a ball $B^3$ from
$\mathbb{R}^3$ and making appropriate identifications on the boundary. For
example, for $\mathbb{R}^3\# T^3$, taking the ball to be a cube, we
identify the
opposite faces of its boundary to obtain $\mathbb{R}^3\# T^3$ as in Figure
(\ref{fig7}).

The fundamental group $\pi_1(\mathbb{R}^3\# T^3)$ is nonabelian. We do not
need
its details here. It is sufficient to know that it has vector bundles with
flat
connections associated to non-abelian UIRR's of this group.

\section{On How Diffeos Can Change Domains}

The reason why this can happen is as follows. Diffeos of a manifold $M$
which
leave a point $P$ (and a frame at $P$) of $M$ fixed act on $\pi_1(M)$. The
reason why we require the diffeos to leave a point $P$ fixed is that
elements of
$\pi_1(M)$ are equivalence classes of loops starting and ending at a
\emph{fixed} point, which we can take to be $P$. One says that the loops
are
thus based at $P$. Diffeos leaving $P$ fixed map loops based at $P$ to
other
loops based at $P$, and hence they act on $\pi_1(M)$ \footnote{Changing the
base point on a pathwise connected manifold gives the ``same'' (that is, isomorphic) group $\pi_1(M)$.}. 

For asymptotically flat spaces such as those we consider, we take $P$ to be
the
``point at $\infty$''. The diffeos of $D^\infty$ become trivial at
$\infty$, and
so leave $P$ fixed. It follows that $D^\infty$ acts on
$\pi_1(\mathbb{R}^d\#
P)$.

But $D^\infty_0$ acts trivially on $\pi_1(\mathbb{R}^d\# P)$. For if $d\in
D^\infty_0$, there is a curve $d_t\in D^\infty_0$ such that $d_1^\infty=d$,
$d_0^\infty=e$. The action of $d$ on a loop $l$ based at $P$ can thus be
continuously deformed to the action of identity. Now the loops
$d_t~l=\langle
d_t x(t): x(0)=x(1)=P \rangle$ are all based at $P$ and as
$t$ decreases from $1$ to $0$, it deforms
$d_1~l=d~l$ to $l$ without ever changing the base point $P$ of the
intermediate
loops. Thus
$d~l$ is homotopic to $l$ and $D^\infty_0$ acts trivially on
$\pi_1(\mathbb{R}^d\# P)$.

But ``large diffeos''  can act nontrivially on $\pi_1(\mathbb{R}^d\# P)$.
This
action is an automorphism as well. Also since $D^\infty_0$ acts trivially,
the
effective action is just that of $D^\infty/D^\infty_0$, the mapping class
group.

For $d\in D^\infty$, let $\tau(d)$ denote this automorphism on
$\pi_1(\mathbb{R}^d\# P)$. If $\rho$ is a representation of
$\pi_1(\mathbb{R}^d\# P)$, set
\begin{equation}
      \left[\tilde{\tau}(d)\rho\right](g)=\rho(\tau(d)^{-1} g),
\end{equation}
for $g\in\pi_1(\mathbb{R}^d\# P)$. Then $\tilde{\tau}(d)\rho$ is a
representation too, which may or may not be equivalent to $\rho$. If the
two
representations are inequivalent, we write
\begin{equation}
      \label{inequiv-rep-1}
      \tilde{\tau}(d)\rho\neq \rho.
\end{equation}

Since $\rho$ fixes the domain, in the case (\ref{inequiv-rep-1}), we can be
sure
that the diffeo $d$ is domain-changing and hence anomalous.

A transformation $\tilde{\tau}(d)$ can be anomalous even if
$\tilde{\tau}(d)\rho=\rho$. We will come back to
this point later.

\subsection{The Diffeo Anomaly for $\mathbb{R}^2\# T^2$}

It is well-known that the mapping class group of $T^2$ is
$SL(2,\mathbb{Z})$.
See \cite{birman1974braids}. This is not quite the same as $D^\infty/D^\infty_0$ of
$(\mathbb{R}^2\#
T^2)$. The latter is the Steinberg group $St(2,\mathbb{Z})$
\cite{Aneziris:1989cr}.

We can describe $St(2,\mathbb{Z})$ as follows. $SL(2,\mathbb{R})$ is
infinitely
connected. Call its universal covering group $\widetilde{SL}(2,\mathbb{R})$ ,
so
that $\widetilde{SL}(2,\mathbb{R})/ \mathbb{Z}=SL(2,\mathbb{R})$. The
inverse image of $SL(2,\mathbb{Z})$ in $\widetilde{SL}(2,\mathbb{R})$ is
$St(2,\mathbb{Z})$, so that $SL(2,\mathbb{Z})=St(2,\mathbb{Z})/\mathbb{Z}$.
The
$\mathbb{Z}$ here corresponds to  ``$2\pi N$-rotations at $\infty$''.

The Steinberg group $St(2,\mathbb{Z})$  can be presented as follows:
\begin{equation}
      \label{st2Z-presentation}
      St(2,\mathbb{Z})=\langle e_{12},e_{21},d:~d=(e_{12} e_{21}^{-1}
e_{12})^4,
de_{ij}=e_{ij}d\rangle.
\end{equation}
Thus $d$ generates the center $\mathcal{C}$ of $St(2,\mathbb{Z})$. It
generates
``$2\pi$-rotations at $\infty$''. The group $SL(2,\mathbb{Z})$ is
$St(2,\mathbb{Z})/\mathcal{C}$.

We want to examine the action of $St(2,\mathbb{Z})$ on
$\pi_1(\mathbb{R}^2\#
T^2)$. For this, by (\ref{st2Z-presentation}), it is enough to show the
action
of $e_{ij}$ on the $a$ and $b$ cycles.

\begin{figure}
  \centering
  \subfloat[The $a$ cycle]{\label{fig:gull-1}\includegraphics[scale=0.25,angle=-90]{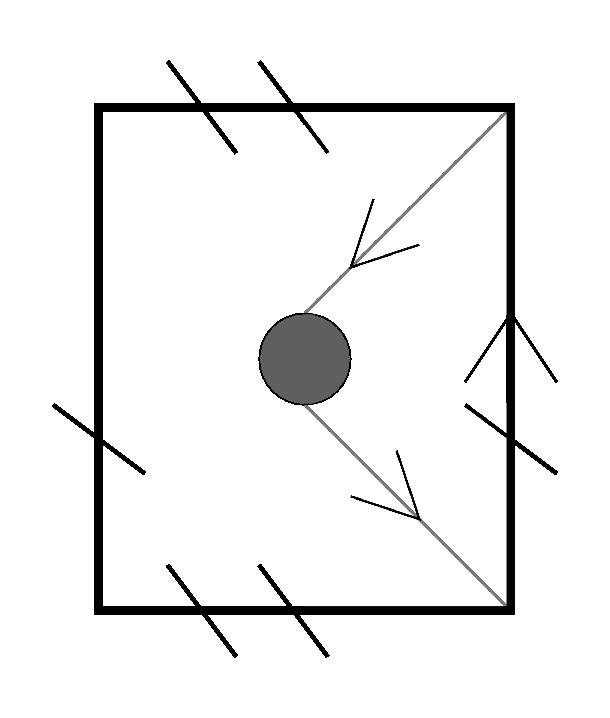} }                
  \subfloat[The $b$
cycle]{\label{fig:tiger-2}\includegraphics[scale=0.25,angle=-90]{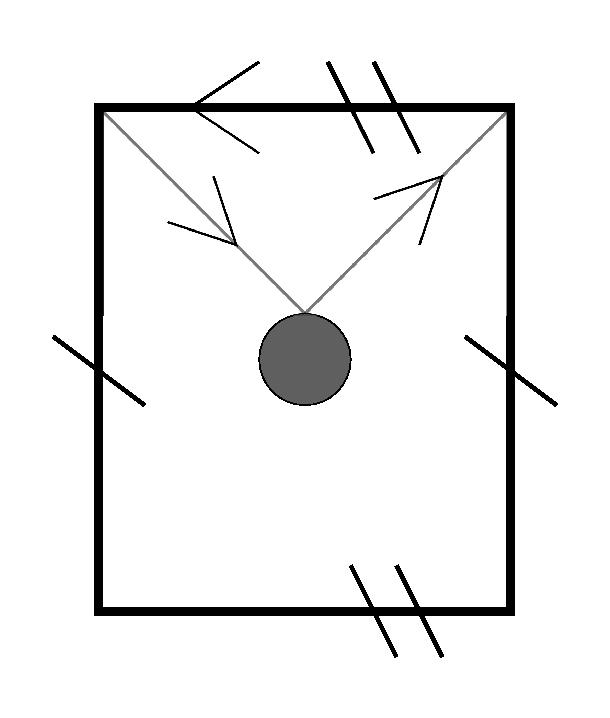}}
   \caption{The $a$ and $b$ cycles of $\mathbb{R}^2\# T^2$ with infinity as
base
point.}
  \label{fig8}
\end{figure}

In Figure \ref{fig6}, the base point is not at $\infty$. So choosing it at
$\infty$, the above cycles are renamed in Figure \ref{fig8}: the
equivalence
class of the cycle $a$ is represented by $N\times N$ matrix $U_1$ and that
of the cycle
$b$ by the $N\times N$ matrix $U_2$.

Under $e_{ij}$, one has the following\footnote{The details of these results
follow from \cite{Aneziris:1989cr}.}:
\begin{eqnarray}
      e_{12}U_1=U_1, &  & e_{12}U_2=U_2U_1, \\
      e_{21}U_1=U_1U_2, & &  e_{21}U_2=U_2.
\end{eqnarray}
As $e_{ij}$ are automorphisms, these equations imply that
\begin{eqnarray}
      e_{12}U_1^{-1}=U_1^{-1}, &  & e_{12}U_2^{-1}=U_1^{-1}U_2^{-1}, \\
      e_{21}U_1^{-1}=U_2^{-1}U_1^{-1}, & &  e_{21}U_2^{-1}=U_2^{-1}.
\end{eqnarray}
The action on the center $c$ in (\ref{root-unity-c}) also follows from the automorphism property:
\begin{equation}
      e_{ij}c=e_{ij}(U_1)e_{ij}(U_2)e_{ij}(U_1^{-1})e_{ij}(U_2^{-1})=c
\end{equation}
So $c$ is invariant under $St(2,\mathbb{Z})$.

But that is \emph{not} the case with $U_1^N$ and $U_2^N$. We find 
\begin{align}
      e_{12} U_1^N &= U_1^N \\
      e_{12} U_2^N &= U_2U_1U_2U_1...U_2U_1 = c^{\frac{N(N-1)}{2}} U_2^N
U_1^N=c^{\frac{N(N-1)}{2}} e^{i N \theta_1} e^{i N \theta_2} \\
      e_{21} U_1^N &= U_1U_2U_1U_2...U_1U_2 = c^{\frac{N(N+1)}{2}} U_2^N
U_1^N=c^{\frac{N(N+1)}{2}} e^{i N \theta_1} e^{i N \theta_2} \\
      e_{21} U_2^N &= U_2^N.
\end{align}

Thus after the diffeo $e_{12}$, $U_1^N$ has the representation
$e_{12}
U_1^N=U_1'^N=U_1^N$, while $U_2^N$ has the representations
$e_{12}U_2^N=U_2'^N=e^{\frac{N(N-1)}{2}}e^{iN\theta_1}e^{iN\theta_2}$.
Since in general $U_2'^N\neq U_2^N$, $e_{12}$ in general changes the UIRR
of
$\pi_1(\mathbb{R}^2\# T^2)$.

Hence the diffeo $e_{12}$ is anomalous for a generic quantization of
$\mathbb{R}^2\# T^2$.

Similar remarks are valid for $e_{21}$.

\subsection{Remarks}

\begin{enumerate}
      \item We can be certain that $e_{ij}$ are domain-changing operators
and
therefore anomalous when they alter the UIRR of $\pi_1(\mathbb{R}^2\# T^2)$
on
which a quantum theory is based.
      
      For $c=1$, $U_i$'s are phases. Then $e_{ij}$ are anomalous if they
change
$U_1$ or $U_2$: that will change the domain.
      
      A simple example is the following:
	    \begin{equation}
	          U_1=e^{i\theta_1}, ~~ U_2=e^{i\theta_2}, ~~ c=1.
	    \end{equation}
	    Then
	    \begin{equation}
	          e_{12}U_1=e^{i\theta_1}, ~~~
e_{12}U_2=e^{i(\theta_1+\theta_2)}.
	    \end{equation}
      So $e_{12}$ is anomalous if $e^{i\theta_1}\neq 1$.

      \item Suppose $c\neq \mathbb{1}$ so that $U_1U_2\neq U_2U_1$, meaning
that
$U_i$'s \emph{themselves} are anomalous. That means that we \emph{cannot}
implement $U_i$'s
 as operators leaving the domain of $i\slashed{D}$ invariant. That is
because
they can be thought of as acting by conjugation or adjoint action
$\textrm{Ad}~U_i$ on the $U_i$'s, that is,
 \begin{equation}
       \textrm{Ad}~U_i (U_j)=U_i U_j U_i^{-1}.
 \end{equation}
But RHS$~\neq U_j$ if $i\neq j$. Thus although the UIRR $\rho$ is
invariant, the
boundary conditions and hence the domain are not and $U_i$'s are anomalous.

Another way to say this is as follows. The group $\pi_1(\mathbb{R}^2\#
T^2)$ is
a
(discrete) \emph{gauge} group. Only \emph{gauge invariant} objects are
observables. But if $\pi_1(\mathbb{R}^2\# T^2)$ has a nonabelian
representation
$\rho$ in a quantum theory, then only the \emph{center}  of
$\rho\left(\pi_1(\mathbb{R}^2\# T^2) \right)$ commutes with all elements of
$\rho\left(\pi_1(\mathbb{R}^2\# T^2) \right)$. Only they are observable.
This center is in general spanned by $c,U_1^N, U_2^N$.
\end{enumerate}

These remarks generalize to \emph{any} nonabelian gauge group and is the
basis of the proof that nonabelian monopoles break color \cite{Balachandran1983,Balachandran1984,Balachandran1984b}.

\section{Witten Spinors}

There is an aspect of the preceding discussion which requires elaboration.
The
spatial Dirac (or Laplace) operator is written in a background metric
outside
the cut-out balls. This metric is flat, the complement of the cut-out ball
being
the asymptotic region. The entire effect of the mapping class group is
treated
in terms of its action on domains of $i\slashed{D}$. It would be attractive
to
derive an approach where the metric away from the geon is not necessarily
flat.

One possibility is to use the Witten spinor $\xi$ or its generalizations
\cite{Witten1981,Gibbons1983enerpos}. In Witten's original work, they were used to prove the positivity
of
the ADM energy in asymptotically flat space-times. The spinor $\xi$ obeys
the
two-component Dirac equation
\begin{equation}
      \label{witten-weyl}
      i\sigma\cdot D \xi = 0,
\end{equation}
where $D_i$ here comes from the four-dimensional covariant derivative
$D_\mu$ by
restricting $\mu$ to spatial indexes. Also $\xi$ approaches a constant
$\xi_0$
(that is to say, a covariantly constant $\xi_0$ with respect to a flat
connection)
at $\infty$. The physical meaning of $\xi_0$ is that $\xi_0^\dagger\sigma_\mu
\xi_0$, with
$\sigma_0=\mathbb{1}$ and $\sigma_i=$ Pauli matrices, determines a
future-pointing null vector. That means that it goes towards fixing a
global
time direction.

One can also work with Dirac spinors instead,
\begin{equation}
      \label{witten-dirac}
i\gamma \cdot D \xi =0,
\end{equation}
with $\xi\to\xi_0=a$ constant Dirac spinor as $|\vec{x}|\to\infty$, so that
$\bar{\xi}\gamma\xi$ is a future-pointing time-like vector. There are
advantages
in using Dirac spinors since they will eliminate the need for
distinguishing
dotted and undotted spinors.

The positive energy theorem solves (\ref{witten-weyl}) and
(\ref{witten-dirac})
for a specified $\xi$
at $\infty$ using the Green's function of $i\sigma\cdot D$ or $i\gamma\cdot
D$
regarded as a self-adjoint operator with Dirichlet boundary conditions  at
$\infty$. That specifies its domain.

By adding source terms to the Einstein tensor $G_{\mu \nu}$, Witten's proof
can
be modified to include matter as well. We can also include nonabelian gauge
fields in
$D$ and carry the proof through after a modification of the dominant energy
condition.

Gibbons \emph{et al.} \cite{Gibbons1983enerpos} have generalized Witten's proof in another
direction by considering black holes and treating the event horizon as a
boundary. That involves a choice of boundary conditions at the horizon which
is
compatible with ellipticity and self-adjointness and is non-trivial.

The point of these remarks is the following. In $(3+1)$ dimensions, we can
attach a geon as previously, that is, by carving out a ball $B^3$ from
$\mathbb{R}^3$ and making identifications. We can then study the operator
$i\sigma\cdot D$ or $i\gamma\cdot D$ where the metric need be only
asymptotically flat and $D$ can include gauge fields. There is little doubt
then
that mapping class groups will generically change its domains and will be
anomalous. In this approach, the metric away from $B^3$ is not constrained
to be flat: it need be only asymptotically flat. 

The positive energy proof however requires that there is a \emph{unique}
extension of the boundary data to the bulk. That requires that the Pauli or
Dirac operators regarded as self-adjoint operators (with domains speciefied
as
in previous sections) have no kernel. We have not tried a proof of this
point,
which is tied up with issues of ellipticity. 

\section{On Black Hole Entropy}

We continue to focus on $\mathbb{R}^2\# T^2$.

Although we begin with a domain $\mathcal{D}_{U_1,U_2}$ fixed by the elements
$U_i$ of the rational torus, the mapping class group generates an orbit in
the space of domains.

Suppose that the dimension of the orbit $D^\infty/D_0^\infty~\mathcal{D}_{U_1U_2}$ is $d(U_1,U_2)$. Then if $|\psi_{U_1,U_2}\rangle\langle \psi_{U_1,U_2}|$ is a state where
$|\psi_{U_1,U_2}\rangle\in \mathcal{D}_{U_1,U_2}$, restoration of diffeo
invariance requires that we average this state over orbit $D^\infty/D_0^\infty~\mathcal{D}_{U_1U_2}$ and work with
\begin{equation}
      \omega=\frac{\Omega}{\Tr \Omega},
\end{equation}
where
\begin{equation}
      \Omega=\sum_{U'_1,U'_2\in D^\infty/D_0^\infty~\mathcal{D}_{U_1U_2}} 
|\psi_{U'_1,U'_2}\rangle\langle
\psi_{U'_1,U'_2}|.
\end{equation}
The state $\Omega$ is not pure. Its entropy is very roughly
\begin{equation}
      S=- \log d(U_1,U_2).
\end{equation}
Entropy can thus be generated from attempts at restoration of symmetries.

There are classical singularities theorems of Friedman, Schleich, Witt \cite{Friedman1993}, Gannon \cite{gannon1975singularities,gannon1976topology} showing that if the spatial slice in $3$d or $4d$ is not
$\mathbb{R}^2$ or $\mathbb{R}^3$, then the space-time obtained from initial
data
is geodesically  incomplete. This theorem is often interpreted to mean that
the
above data evolve into black holes. Geons are hence classically expected to
evolve into black holes. The entropy $S$ may then contribute to the entropy of the
black
hole.

This is an interesting possibility, but to make connection to the usual
black
hole thermodynamics, $S$ must scale with the area of the horizon. We do not obtain this
scaling in any obvious way.

\section{Acknowledgement}

APB and ARQ would like to thank Prof. Alvaro Ferraz (IIP-UFRN-Brazil) for
the
hospitality at IIP-Natal-Brazil, where part of this work was carried out.
APB
also acknowledges CAPES for the financial support during his stay at
IIP-Natal-Brazil. APB is
supported by DOE under grant number DE-FG02-85ER40231. ARQ is supported by
CNPq
under process number 307760/2009-0.


\end{document}